# High-efficiency fast pinching radiation of electron beams in nonuniform plasma

**Authors:**

Xing-Long Zhu[1*], Min Chen[2,3], and Zheng-Ming Sheng[2,3,4*]

**Affiliations:**

[1] Institute for Fusion Theory and Simulation, School of Physics, Zhejiang University, Hangzhou 310058, China

[2] Key Laboratory for Laser Plasmas (MOE), School of Physics and Astronomy, Shanghai Jiao Tong University, Shanghai 200240, China

[3] Collaborative Innovation Center of IFSA, Shanghai Jiao Tong University, Shanghai 200240, China

[4] Tsung-Dao Lee Institute, Shanghai Jiao Tong University, Shanghai 201210, China

*Email: xinglong.zhu@zju.edu.cn (X.L.Z.); zmsheng@sjtu.edu.cn (Z.M.S.)

**Abstract:** The continuous development of bright x/gamma-ray sources has opened up new frontiers of science and advanced applications. Currently, there is still a lack of efficient approaches to produce gamma-rays with photon energies up to GeV and with high peak brilliance comparable to modern free-electron lasers. Here we report a novel mechanism called beam fast pinching radiation burst to generate such gamma-ray sources. It is achieved by injecting a GeV electron beam into a submillimeter plasma with an upramp density profile, enabling violent beam pinching to occur rapidly. During this process, a burst of collimated gamma-rays is efficiently produced with photon energy up to GeV, energy conversion efficiency exceeding 30%, and peak brilliance exceeding $10^{28}$ photons s$^{-1}$ mm$^{-2}$ mrad$^{-2}$ per 0.1% bandwidth. All of these are several orders of magnitude higher than existing gamma-ray sources. This opens a novel avenue for the development of extremely bright gamma-ray sources for both fundamental research and cutting-edge applications.

Bright x-ray sources from synchrotrons or x-ray free-electron lasers (XFELs) have become crucial tools to advance science and applications in broad areas [1], including medicine, materials science, and industry. Currently, these x-ray sources are mostly based upon large-scale radio-frequency accelerators, making them available only in a few national laboratories. Moreover, the photon energy from these sources is generally limited to the range of a few keV to hundreds of keV. There is a huge demand and significant interest in developing compact radiation sources and extending photon energy to well beyond MeV with high efficiency and high brilliance, as they have unique applications in basic science, such as exploring strong-field quantum electrodynamics (QED) physics [2, 3], probing nuclear structure and reaction dynamics [4, 5], developing gamma-ray photon colliders [6], and enabling frontier research in relativistic astrophysics from gamma-ray bursts to the formation of lepton-dominated jets [7, 8].

In addition to these based upon the conventional accelerator technologies, plasma-based accelerators driven by either intense laser pulses or high-energy particle beams are being developed as compact accelerators and radiation sources [9-13]. A nonlinear plasma wake can support large acceleration gradients ~100GV/m, which are more than three orders of magnitude higher than those found in conventional accelerators. The x-ray emission is produced either via betatron radiation [14-18], nonlinear Thomson, or inverse Compton scattering [19-21]. For betatron radiation, considerable efforts have been made to enhance their photon energy and/or brilliance, for example, by use of resonant betatron oscillations [22], a passive plasma lens [23], and density tailored plasmas [24, 25]. They usually have limited field strength and low beam density. So far, the energy of emitted photons measured in experiments is limited to the level of a few keV to hundreds of keV, the peak brilliance is limited to the order of $10^{23}$ photons s$^{-1}$ mm$^{-2}$ mrad$^{-2}$ per 0.1% bandwidth (BW), and the energy conversion efficiency is generally on the order of 0.001%. For nonlinear Thomson and inverse Compton scattering, the peak brilliance and the energy conversion efficiency are also relatively low and the experimental implementation is usually challenging as the requirements for spatial-temporal control between the electron beams and laser beams are very high. Nevertheless, this method of laser-electron scattering has the potential to produce gamma-rays above MeV, for example, the planned gamma-ray facility at the Extreme Light Infrastructure for Nuclear Physics [5] is expected to deliver gamma-ray pulses with photon energy up to 20MeV and peak brilliance in the range of $10^{20}$–$10^{23}$ photons s$^{-1}$ mm$^{-2}$ mrad$^{-2}$ per 0.1% BW. Gamma-rays can also be produced by bremsstrahlung of energetic electrons [26, 27] or proposed methods with the next generation of ultrahigh-intensity lasers [28-32]. However, these sources are normally subject to large divergence and large size, severely limiting their brilliance and efficiency. Overall, despite significant advances made so far, it is still a huge challenge to efficiently generate extremely bright gamma-ray sources with photon energies up to GeV and with high peak brilliance comparable to modern XFELs. To achieve this, not only is an efficient radiation mechanism needed to convert ultra-relativistic electrons into high-energy gamma-rays, but also the resulting beam needs to be tightly focused with high density and low divergence. How to meet these requirements simultaneously remains elusive.

In this Letter, we report a promising and unique approach called the beam fast pinching radiation burst (FPRB), which can dramatically enhance the gamma-ray brilliance, photon energy, and energy conversion efficiency. In our approach, an ultra-relativistic electron beam is adopted to drive plasma wakefields in nonuniform plasma with an upramp density profile, where the electron beam is self-pinched quickly as a whole during the beam propagation in plasma. This causes the highly nonlinear wake in the dense plasma region to be further enhanced both longitudinally and transversally to ultrahigh field strengths up to tens of TV/m. Consequently, intense high-energy gamma-ray emission is effectively triggered with an efficiency several orders of magnitude higher than existing gamma-ray sources, and in particular, the resulting gamma-ray brilliance can be increased to near levels of modern XFELs and the photon energy can be extended into the GeV range. Such powerful gamma-rays could provide new capabilities for fundamental research and applications.

The proposed approach for efficient generation of brilliant GeV gamma-rays is illustrated schematically in Fig. 1. When an ultra-relativistic electron beam passes through a plasma with an upramp density profile along its propagation direction, a nonlinear plasma wake is excited with a plasma cavity structure around the beam, which comoves with the beam. The transverse fields inside the cavity tend to focus the beam during its propagation. As the beam is pinched along the propagation distance, the plasma density is increased correspondingly with the given nonuniform plasma density profile. This, in turn, can drive stronger wake fields in a high-density plasma with ultrahigh amplitude up to tens of TV/m, enabling a novel regime of beam-driven high-energy radiation. Obviously, this is completely different from traditional betatron radiation [14-18, 22-25]. An effective energy conversion mechanism from the electron beam to photon emission can be found, where a highly dense GeV gamma-ray pulse with a submicron spot size is obtained, resulting in unprecedented levels of radiation efficiency, photon energy, and flux. As a consequence, the peak brilliance of the emitted gamma-rays can be increased by several orders of magnitude over current gamma-ray sources.

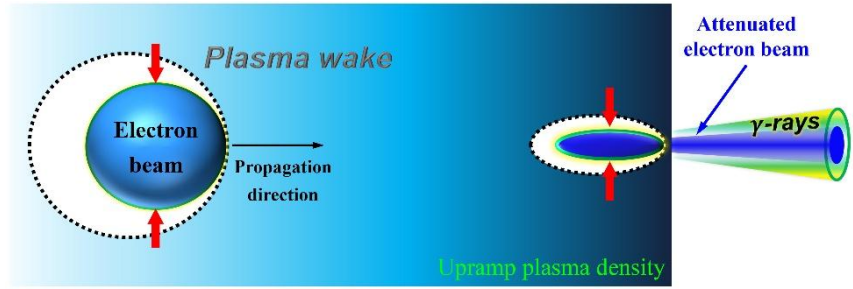

**Fig. 1.** Schematic illustration. When an ultra-relativistic electron beam is incident into nonuniform plasma with an upramp density profile, strong beam pinching occurs due to the rapid excitation of the enhanced plasma wakefields with a cavity structure around it. The focused dense electron beam can, in turn, drive stronger wakefields up to tens of TV/m in a high-density plasma region, thereby leading to a burst of gamma-ray emission with extremely high efficiency and brilliance. The red arrows indicate the direction of the pinching force experienced by the electron beam.

To demonstrate the proposed scheme, three-dimensional (3D) particle-in-cell (PIC) simulations were carried out using the code EPOCH [33]. As an example, we take a drive electron beam with velocity $v$ moving along the $x$-axis, which has about 1nC charge, 5GeV mean energy, 5 mm-mrad normalized emittance, energy spread of 5% FWHM, and Gaussian density distribution of $n_b = n_{b0} \exp\left(-\frac{r^2}{\sigma_{r0}^2} - \frac{(x-vt)^2}{\sigma_{x0}^2}\right)$, where $\sigma_{x0} = 2$μm, $\sigma_{r0} = 2.5$μm, and $n_{b0} = 2 \times 10^{26}$m$^{-3}$. These parameters are within the reach of existing acceleration techniques [34-36]. For the parameters considered, the beam self-current is more than three orders of magnitude lower than the Alfvén current limit $I_A = m_e \gamma_b c^3 / e \approx 17\gamma_b$ kA, where $m_e$ is the rest electron mass, $e$ is the elementary charge, $\gamma_b$ is the relativistic Lorentz factor, and $c$ is the speed of light in vacuum. A nonuniform plasma slab with an upramp density profile is used as the conversion target to realize strong self-pinching of the electron beam and subsequent effective emission of high-energy photons, where the plasma target has a linearly increasing density profile $n_p = n_{p0}(x/L)$ ranging from 0 to $n_{p0} = 5 \times 10^{27} \text{m}^{-3}$ over the longitudinal distance of $L = 400\mu\text{m}$. Note that this linear density profile is adopted as an example for simplicity, and our scheme also works for other plasma profiles with positive density gradients, as discussed in the Supplemental Material [37] (including Refs. [38-43]). Both beam and target parameters are tunable. A simulation window moving along the $x$ direction with velocity $c$ is employed, which has a size of $8\mu\text{m}(x) \times 6\mu\text{m}(y) \times 6\mu\text{m}(z)$ with $400 \times 300 \times 300$ grid cells. The macroparticles in each cell for the beam electrons and plasma electrons (ions) are 27 and 8 (8), respectively. Note that the PIC approach is capable of self-consistent simulation of electron beam-plasma interactions [44-47].

Figure 2 shows the simulation results for strong beam focusing and dense gamma-ray emission. Evolution of the energy trajectory of selected beam electrons is shown in Fig. 2(a). As it shows, the beam radius decreases along the propagation distance quickly, so its density increase significantly according to $n_b \propto n_{b0}\sigma_{r0}^2/r^2$ [Fig. 2(b) compared with Fig. 2(e)]. Such a focused beam has the ability to drive much higher fields in denser plasma, so that ultrahigh fields in excess of $10^{13}$ V/m are excited [see Figs. 2(c) and 2(f)]. It is difficult for traditional beam-driven plasma wakes to realize such a high-strength field. Currently, it is usually proposed to reach such fields by use of ultraintense laser fields in the laser-electron collision configurations [48]. However, the radiation efficiency with such configurations is typically very low, making it difficult to produce ultrabright gamma-ray sources, let alone the technical challenges associated with shot-to-shot fluctuations and very demanding alignment and synchronization requirements between the laser and electron beams. In such strong fields, the QED parameter $\chi_e = \left(\frac{e\hbar}{m_e^3 c^5}\right) \varepsilon_b |\mathbf{E}_{\text{eff}}|$ can exceed 0.2, allowing the radiation reaction effect to play a considerable role in photon emission, which may extend radiation emission of high-energy electrons in plasma wakefields into a new parameter regime. Here $\hbar$ is the reduced Planck constant, $\varepsilon_b = \gamma_b m_e c^2$ is the electron energy, $\mathbf{E}_{\text{eff}} = (\mathbf{E}_\perp + \boldsymbol{\beta} \times \mathbf{B})$ is the effective interaction field experienced by the electron, and $\mathbf{E}_\perp$ is the electric field term perpendicular to the normalized electron velocity $\boldsymbol{\beta} = \mathbf{v}/c$. As a consequence, each radiated photon can gain a large fraction of the electron energy, which can reach more than 1 GeV. Thanks to the ultrahigh density and submicron diameter of the focused electron beam, the resulting gamma-ray pulse has a similarly small size and high density [see Fig. 2(g)], which gives extremely high brilliance.

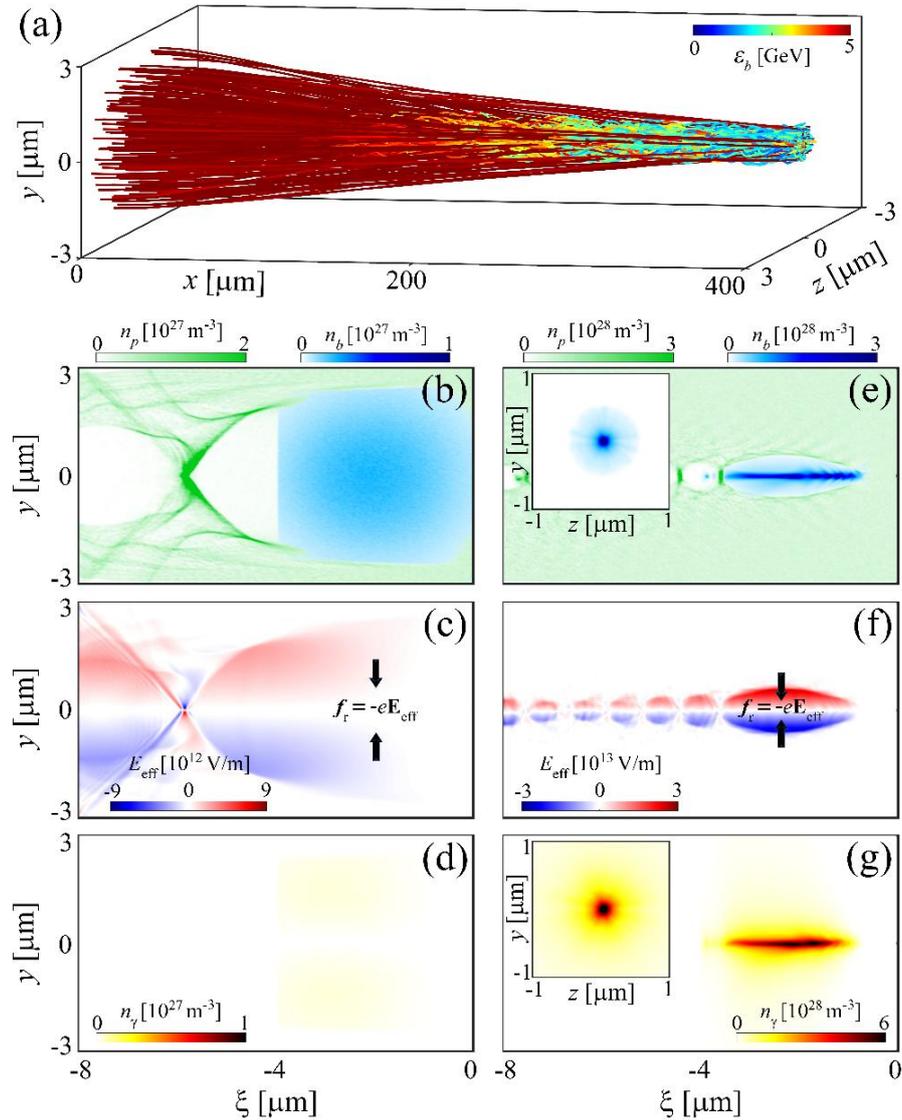

**Fig. 2.** (**a**) Trajectories of selected electrons in the beam and their energy evolution in 3D geometry. Distributions of (**b**, **e**) the plasma density ($n_p$) and the beam density ($n_b$), (**c**, **f**) the effective interaction field ($E_{\text{eff}}$) and (**d**, **g**) the photon density ($n_\gamma$) are shown at the beginning (left column: **b** to **d**) and the end (right column: **e** to **g**) of the plasma target. The inset in (**e**) exhibits the transverse distribution of the final electron beam. The inset in (**g**) displays the transverse distribution of the final gamma-ray beam.

6 / 12

Evolution of the energy spectrum of the electron beam and gamma-rays is shown in Fig. 3. After interaction with the plasma with a density scale length of 400μm, the energy conversion efficiency of the electron beam to gamma-rays is as high as nearly 31%, with photon energies up to GeV. Due to the strong energy dissipation caused by high-energy photon emission, the electron beam energy spectrum is modified radically, changing from a single peak of 5GeV to a broad spectrum with two peaks of approximately 1.6GeV and 5GeV due to significant energy transfer from the electron beam to the gamma-rays. The gamma-ray source has a small divergence of approximately 3mrad FWHM. Assuming a source size of 0.2μm, a duration of 5fs and $1.8 \times 10^{10}$ photons, one can estimate the peak brilliance to be $1 \times 10^{28}$ photons s$^{-1}$ mm$^{-2}$ mrad$^{-2}$ per 0.1% bandwidth at 1MeV, which is approximately five orders of magnitude higher than current gamma-ray sources. Note that as the gamma-rays are produced simultaneously during the electron beam pinching process and they co-propagate with the beam, the gamma-ray pulse should have a duration comparable to the electron beam duration.

Physically, the rapid pinching of the electron beam in the upramp density plasma is responsible for the produced radiation. To understand this, we have developed a theory model, the details of which are given in the Supplemental Material [37]. According to the model, it is found the beam radius $r_f$ changes along the propagation direction with

$$r_f/r_0 \approx 0.9 S^{-1/4} \cos\left(\frac{2}{3}S^{3/2} - \frac{\pi}{12}\right), \tag{1}$$

where $r_0$ is the initial beam radius and $S = \left(\omega_{p0}^2/2\gamma_b L c^2\right)^{1/3} x$ with $\omega_{p0}^2 = 4\pi n_{p0} e^2/m_e$, $x$ the propagation distance, and $L$ the plasma density scale length, obtained under $S \gg 1$. This can be further approximated as $\bar{r}_f/r_0 \sim 0.5 S^{-1/4}$ after averaging $|\cos\theta|$ over the range $[0, 2\pi]$. This model indicates clearly that a proper upramp density gradient is the key for the rapid beam focusing, which finally leads to efficient gamma-ray emission.

As a comparison, we also investigate the interaction of the electron beam with a uniform density plasma, see the Supplemental Material [37] for more details. It is clear from Fig. 3(b) that the photon energy and yield of the emitted gamma-rays are much lower than these obtained in the plasma with the upramp density profile. The corresponding energy conversion efficiency from the electron beam to gamma-rays is as low as only about 0.3%. Due to the lack of beam focusing, the gamma-ray source has a quite large size comparable to that of the initial electron beam. Consequently, the source brilliance is reduced by approximately four orders of magnitude as compared to the case with the upramp density profile.

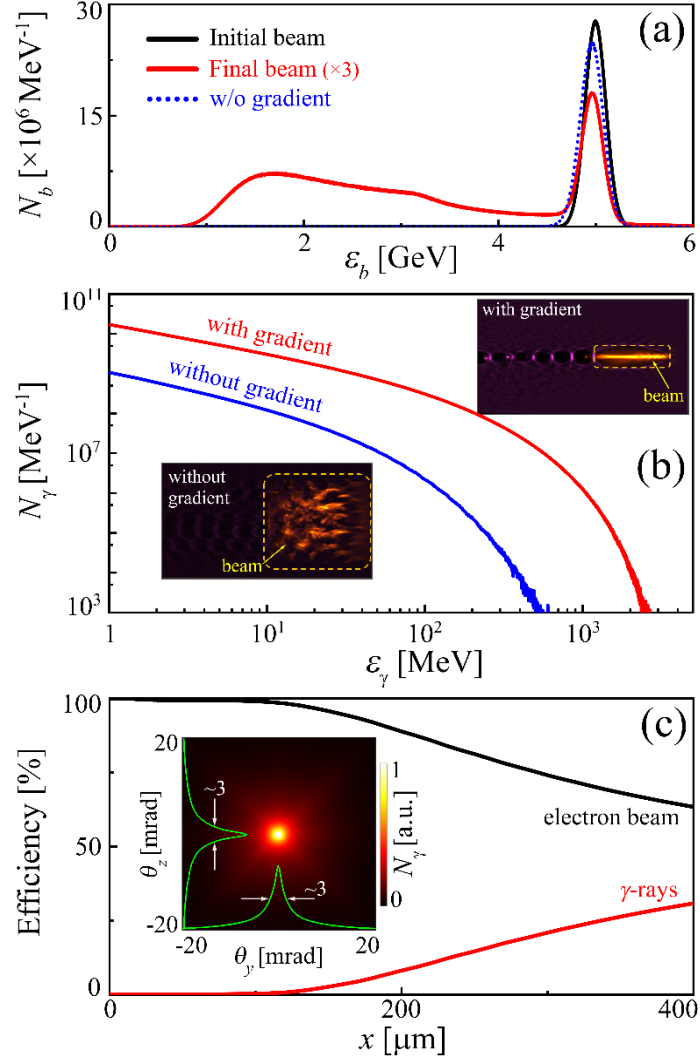

**Fig. 3**. (**a**) The initial (back line) and final (red line) energy spectrum of the electron beam, where the blue dashed line depicts the final beam energy spectrum without a positive density gradient. (**b**) The energy spectrum of the emitted gamma-rays with (red line) and without (blue line) a positive density gradient, where the corresponding insets exhibit the beam focusing in inhomogeneous and homogeneous plasmas, respectively. (**c**) Evolution of the energy efficiency of the electron beam (black line) and gamma-rays (red line), where the inset exhibits the angular distribution of the gamma-ray source.

We now discuss the robustness of this scheme in terms of the beam and plasma parameters. We first investigate the effect of the maximum plasma density (equivalent to density gradient shift) on photon emission, while keeping other parameters fixed. It is shown that a relatively high-density plasma facilitates the electron beam focusing and the excited field enhancement, leading to more efficient photon emission and higher brilliance, as seen in Fig. 4(a). Nevertheless, in order for the entire electron beam to be well confined and focused in the wake, the maximum plasma density should be within certain upper limit given by $n_{p0} < n_{b0} L \exp(-2)/\sigma_{x0} \approx$ $5.4 \times 10^{27} \mathrm{m}^{-3}$, see the Supplemental Material [37] for more details.

Figure 4(b) illustrates the effect of the initial beam density on the gamma-ray radiation. As expected, the driving beam with higher initial density can significantly boost the photon emission, because it can drive denser plasma and so excite stronger wake fields, which gives a larger QED parameter $\chi_e$. Therefore, the radiation efficiency and peak brilliance of emitted gamma-rays can be dramatically enhanced. For example, when $n_{b0} = 4 \times 10^{26} \mathrm{m}^{-3}$, the emitted gamma-rays have an energy efficiency of up to 46% and a high brilliance of about $1.7 \times 10^{28}$ photons s$^{-1}$ mm$^{-2}$ mrad$^{-2}$ per 0.1% bandwidth. It is worth mentioning that, as a large amount of the electron beam energy is lost, the growth in radiation efficiency and peak brilliance gradually slows down both with the further increase of plasma density and beam density.

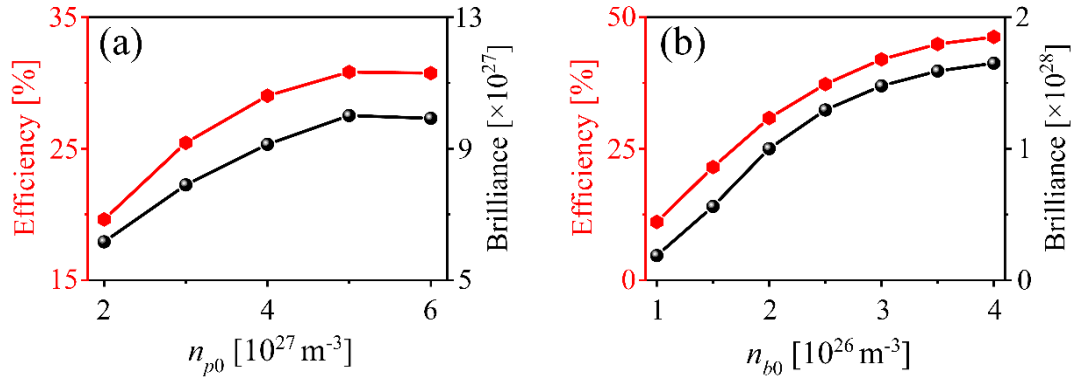

**Fig. 4.** Effects of (**a**) the plasma density ($n_{p0}$), and (**b**) beam density ($n_{b0}$) on the gamma-ray energy efficiency and peak brilliance (photons s$^{-1}$ mm$^{-2}$ mrad$^{-2}$ per 0.1% bandwidth).

We also investigate the effects of the driving electron beam sizes on photon emission in Fig. 5. It is shown that an appropriately large beam is beneficial for high-efficiency generation of bright gamma-rays. If the electron beam has a small radius or short length, it means that its total charge is quite low (where $Q_{b0} \propto n_{b0} \sigma_{r0}^2 \sigma_{x0}$), making it difficult to induce large plasma fields. For example, when $\sigma_{r0} = 1.5 \mathrm{\mu m}$ or $\sigma_{x0} = 1 \mathrm{\mu m}$, $\chi_e$ will be reduced to about 0.1, which causes a significant reduction in the energy efficiency of gamma-rays. On the other hand, the electron beam size should not be too large; Otherwise, it will not be fully confined within the plasma wake, leading to a decrease in the final energy efficiency. Overall, the scheme can be further optimized by changing the size of the electron beam, its density, and the plasma density profile, etc., thus further improving the beam focusing and the resulting photon emission.

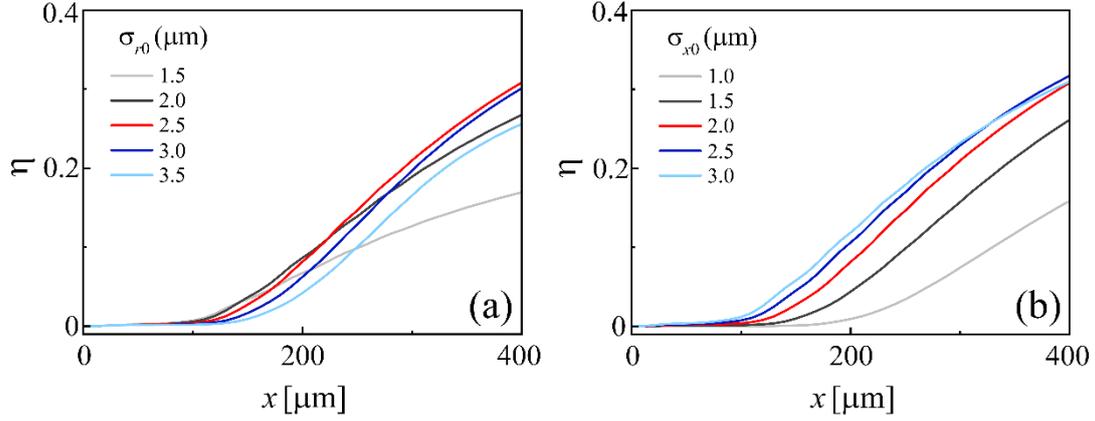

**Fig. 5.** Energy conversion efficiency ($\eta$) of gamma-rays emitted as a function of the interaction distance for (**a**) different transverse beam sizes ($\sigma_{r0}$), and (**b**) different longitudinal beam sizes ($\sigma_{x0}$). Here $\eta$ is defined as the ratio of the gamma-ray pulse energy to the initial electron beam energy.

In summary, we have discovered a high-efficiency radiation mechanism for ultrabright gamma-rays, called FPRB, through a simple configuration of a relativistic electron beam interacting with plasma with an upramp density profile. As the beam travels through this nonuniform plasma, it can be greatly focused to a submicron diameter due to strong beam pinching caused by shrinking plasma wake, increasing its density by hundreds of times. Such a dense beam drives the plasma wake to unprecedented fields of tens of TV/m, enabling one to access the radiation-dominated regime. Under such extreme conditions of interaction, it will radically alter the energy distribution of energetic electrons and their radiated photons. As a result, more than 30% of the beam energy can be converted into intense gamma-ray emission, where the photon energy can be extended to the GeV level. With this highly efficient beam-driven radiation mechanism, the brilliance of the emitted gamma-rays can reach an unprecedented level above $10^{28}$ photons s$^{-1}$ mm$^{-2}$ mrad$^{-2}$ per 0.1% BW, which is about five orders of magnitude higher than existing gamma-ray sources. The involved electron beam parameters may be obtained not only from a conventional accelerator but also from a laser wakefield accelerator as our scheme does not have a high requirement on the beam energy spread, making it highly flexible for experiments. This scheme opens an efficient and promising route for the development of compact bright gamma-ray sources, which is of great significance and interest for both fundamental research and applications [2-8].

**Acknowledgements**

This work was supported by the National Natural Science Foundation of China (Grant Nos. 12205186, 12135009, and 11991074).